\documentclass[aps,prl,twocolumn,groupedaddress,amssymb,superscriptaddress]{revtex4}
\usepackage[dvips]{graphicx}
\usepackage{bm}      % RevTex4 bold math
\newcommand{\ybcoy}{Y\-Ba$_2$\-Cu$_3$\-O$_{6+y}$}
\newcommand{\calabalacuo}{Ca$_{x}$La$_{1.25}$Ba$_{1.75-x}$Cu$_3$O$_{6+y}$}

\newcommand{\ybcocay}{Y$_{1-x}$Ca$_x$\-Ba$_2$\-Cu$_3$\-O$_{6+y}$}

\newcommand{\lsco}{La$_{2-x}$Sr$_x$\-CuO$_4$}
\begin{document}
\bibliographystyle{apsrev}

%Title of paper
\title{Evidence of two distinct charge carriers in underdoped high T$_c$ cuprates.}
\author{S. Sanna}
\email[]{Samuele.Sanna@fis.unipr.it}
\author{F. Coneri} 
\affiliation{Unit\`a CNISM di Parma e Dipartimento di Fisica, I 43100 Parma, Italy}
\author{A. Rigoldi} 
\affiliation{Unit\`a CNISM di Parma e Dipartimento di Fisica, I 43100 Parma, Italy}
\affiliation{Unit\`a CNISM di Cagliari e Dipartimento di Fisica, I 09042 Monserrato (Ca), Italy}
\author{G. Concas}
\affiliation{Unit\`a CNISM di Cagliari e Dipartimento di Fisica, I 09042 Monserrato (Ca), Italy}
\author{R. De Renzi} 
\affiliation{Unit\`a CNISM di Parma e Dipartimento di Fisica, I 43100 Parma, Italy}

\begin{abstract}

We present results on heavily underdoped \ybcocay\ which provide the evidence that the doping mechanism  (cation substitution or oxygen loading) directly determines whether the corresponding injected mobile holes contribute to superconductivity or only to the normal metallic properties. We argue that this hole tagging calls for a subtler description of the correlated bands than the usual one. We also map in great detail the underdoped superconducting phase diagram $T_c$ vs hole doping which shows that the number of mobile holes is not the critical parameter for the superconductivity. 

\end{abstract}

\date{\today}

%\maketitle must follow title, authors, abstract, \pacs, and \keywords
\maketitle

{
%The metal-insulator transition (MIT) in underdoped high $T_c$ cuprates is crucial for developing a model of doped Mott insulators and, eventually, understanding the mechanism of their superconducting behaviour. Doped charges lead to the  formation of a pseudogap, to the appearance of a metallic properties and to superconductivity, but the relation among these three features is far from being resolved. Here we present results on heavily underdoped \ybcocay\ which provide the unprecedented evidence that the doping mechanism  (cation substitution or oxygen loading) directly determines whether the corresponding injected mobile holes contribute to superconductivity or only to the normal metallic properties. We argue that this hole tagging calls for a subtler description of the correlated bands than the usual one, and that our finding is at odds with the interpretation of the pseudogap excitations just as preformed pairs condensing at $T_c$.}

The behaviour of a hole in the cuprate doped Mott-Hubbard insulator is often described in a {\em universal} picture, where, above some critical concentration, it forms the Zhang-Rice singlet \cite{Zhang:1988}, in a single correlated-band scheme. However structural and compositional details of each specific compound do influence the fine grain behaviour. This is particularly true in the region of the metal-insulator (MI)transition , where there is growing evidence that the competition between antiferromagnetic (AF) order and superconductivity is strongly influenced by disorder \cite{Dagotto:2005}, leading to different phase diagrams \cite{Niedermayer:1998,Sanna:2004,Sanna:2007} in different real materials.

Growing evidence that more than one band is needed comes from \calabalacuo, where two distinct charge fluids have been reported \cite{Keren:2006} by NQR. At optimum doping tunneling spectroscopy directly detects \cite{Ngai:2007} two CuO$_2$ gaps in \ybcocay, and $\mu$SR provides additional supporting evidence \cite{Khasanov:2007} in the case of \lsco. In the underdoped regime early NMR \cite{Rigamonti1990,Rigamonti1998} and recent magnetotransport \cite{Ono:2007} results demonstrate additional thermally activated doped holes. The activation energy, proportional to $x^{-1}$, has been shown to scale with relevant ARPES Fermi-arc features and it has been linked directly to the pseudogap \cite{Gorkov:2006}. Further details, such as the presence of Fermi pockets, from high field quantum oscillations \cite{Doiron-Leyraud2007a}, call for a subtler band structure implementation \cite{Elfimov}. Activated holes are also indirectly detected through the magnetic order parameter reduction measured with NQR and $\mu$SR \cite{Borsa:1995,Sanna:2003}. It seems that at least two bands\cite{Sboychakov:2007} are needed to correctly describe real cuprates.

 In order to focus this issue we zoomed into the MI transition region of \ybcocay, where two distinct doping mechanisms can be independently controlled. Charge doping is provided both by heterovalent Ca$^{+2}\rightarrow$Y$^{+3}$ substitution, $x$, and by interstitial oxygen content, $y$, in the Cu(1)O chain layer, yielding a total hole concentration $h=h_{Ca}+h_{O}$ transferred to the active Cu(2)O$_2$ layers. We thus directly show that the two contributions behave very differently with respect to room temperature normal properties and superconductivity.

Polycrystalline samples were prepared by a topotactic technique, which consists in the oxygen equilibration of stoichiometric quantities of the two end members, tightly packed in sealed vessels \cite{Manca:2001}. Low temperature annealing yields high quality homogeneous samples with an absolute error of $\delta x\!\!=\!\!\pm 0.02$ in oxygen content per formula unit (reduced to $\pm 0.01$ after recalibrating end member of different batches). Besides this determination, absolute oxygen content is cross checked by iodometric titration, thermogravimetry on each sample and selected neutron Rietveld refinements. Ca content is checked by X-ray and neutron Rietveld refinements. Reported error bars are the global error from this procedure. Transition temperatures correspond to the linear extrapolation of the 90 \% to 10\% diamagnetic drop of the susceptibility, measured in a field $\mu_0 H = 0.2$ mT. These data agree within quoted uncertainties with resistive determinations (and $\mu$SR, when available). The width of the interval where the resistance drops from 90\% to 10\% of the onset value is typically 6 K for $y\le 0.4$, like for Ca free pellets and large single crystals (see Ref.~\cite{Sanna:2004} and Refs.~therein).

The inset in Fig.~\ref{fig:1} shows the progressive reduction of $T_c$ vs. $x$ in the $y\simeq 0.9$ end members (our starting polycrystals, reacted and annealed in oxygen atmosphere), witnessing the double doping mechanism\cite{Obertelli:1992,Honma:2004}. The superconducting transition measured by SQUID magnetometry is progressively reduced from the maximum $T_c=92K$ at $x=0$ (optimal doping), as Ca substitution injects additional holes:  both $h_{Ca}$ and $h_O$ contribute to superconductivity, driving the samples into the overdoped region. The smooth linear dependence of $T_c(x)$ of the inset in Fig.~\ref{fig:1} also guarantees an effective Ca-Y substitution in the whole explored range. 

However, when tuning of the oxygen content drives these same samples to low doping, close to the MI boundary, we find a markedly different behaviour. Figure \ref{fig:1} displays the critical temperature, $T_c$, versus oxygen content, $y$, for series of samples at fixed calcium content, $0\leq x\leq 0.14$. Surprisingly $T_c$ falls on the same curve for all series with $x\leq 0.09$, giving rise to superconductivity at the same critical oxygen concentration, $y_c=0.30$: the injection of additional holes with calcium does not appreciably influence $T_c$, in our underdoped, $x\le 0.09$ samples. 

%\begin{widetext}
\begin{figure}
\includegraphics[width=0.45\textwidth,angle=0]{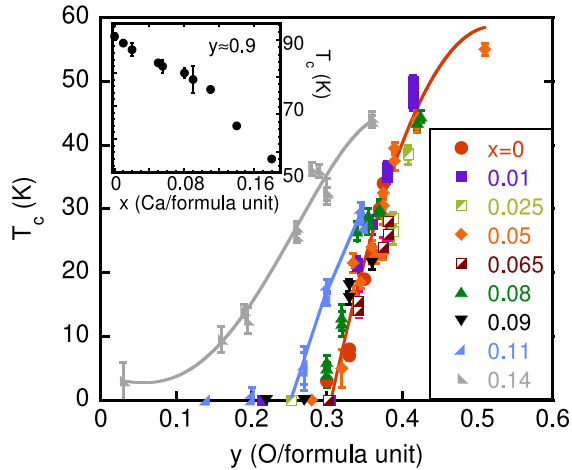}
 \caption[ ]{ Superconducting transition temperatures $T_c$ of \ybcocay\ samples vs.~oxygen concentration for the whole set of samples: The $y\le 0.09$ data fall on the same line (curves are guides to the eye). Inset: transitions $T_c$ detected by zero field cooling SQUID magnetization vs.~calcium concentration, for the fully oxygenated, $y\approx 0.9$ samples: The reduction of $T_c$ with increasing $y$ is the signature of overdoping due to Ca substitution. }

 \label{fig:1}
\end{figure}
%\end{widetext}

Why are holes transferred by Ca and O not additive for superconductivity in the low doping regime? 
The different cationic radii ($R_{Ca^{+2}}/R_{Y^{+3}}\approx 1.1$) may trivially alter oxygen order in the Cu(1)O chains, reducing their charge transfer efficiency. This is the case, e.g., for substantial Y substitutions \cite{Lutgemeier:1996}, where the formation of chains takes place at much larger oxygen content. An independent assessment of the hole content is required to rule out this effect.
The Seebeck coefficient is an independent  measure of the mobile carrier content, since an exponential dependence of $S$ vs $h$ is observed in \ybcoy\ in a large range of doping. \cite{Obertelli:1992,Honma:2004} We systematically measured the value of $S$ at $T=$290 K (RT) in our samples, calibrating the dependence on the fully reduced compounds, $y\approx 0$ samples equivalent to $h_{O}=0$, by assuming an average hole content per Cu plane $h=h_{Ca}=x/2$. We identify two regions, as in previous work \cite{Obertelli:1992}, and our best fit to  $S(h)=\alpha exp(-\beta h)$ shown in the inset of Fig.~\ref{fig:2}, yields values $\alpha=480\,\mu$V/K and $\beta=25$ for $h>0.016$, $\alpha=650\,\mu$V/K and $\beta=44$ for $h<0.016$.
 
%Thermopower calibration, performed at $T=290$ K on $y\approx 0$ samples, equivalent to $h_{O}=0$, assume an average hole content per Cu plane $h=h_{Ca}=x/2$. Notice that in the entire range of oxygen content explored in this letter, $0.05\le y\le 0.42$, the Cu(1)O chains do not produce \cite{Bernhard:1996}  a normal metal contribution to the thermopower. We identify two regions, as in previous work \cite{Obertelli:1992}, and our best fit to  $S(h)=\alpha exp(-\beta h)$ shown in the inset of Fig.~\ref{fig:2}, yields values $\alpha=480\,\mu$V/K and $\beta=25$ for $h>0.016$, $\alpha=650\,\mu$V/K and $\beta=44$ for $h<0.016$. This is a phenomenological fit and our conclusions are independent of its details. In particular Fig.~\ref{fig:2} is self similar if one assumes a uniform transfer rate $h=kx$, with a different $k$ coefficient. 

Figure~\ref{fig:2} shows the $h$ values obtained for all our \ybcocay\ samples ($0 \leq x \leq 0.14$) by comparing their RT Seebeck coefficient $S$ with the calibration curve of the inset in Fig.~\ref{fig:2}, under the assumption\cite{Obertelli:1992,Honma:2004} of nearly equal mobilities for the two types of holes, $h_O$ and $h_{Ca}$. \footnote{The calibration curve in the inset in Fig.~\ref{fig:2} is a phenomenological fit and our conclusions are independent of its details. In particular Fig.~\ref{fig:2} is self similar if one assumes a uniform transfer rate $h=kx$, with a different $k$ coefficient.}

%\begin{widetext}
\begin{figure}
\includegraphics[width=0.45\textwidth,angle=0]{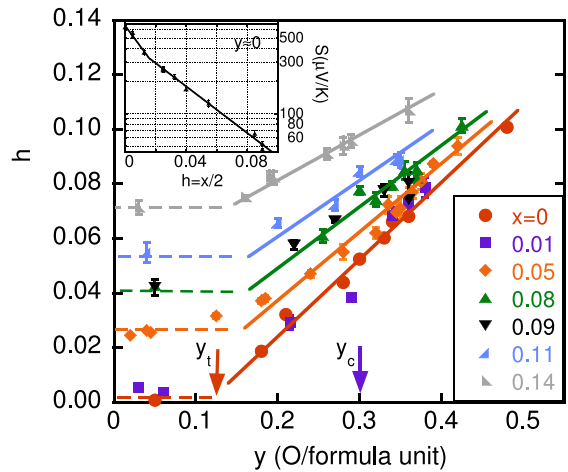} 
\caption[ ]{ Total hole content of \ybcocay\ samples, $h$, obtained from thermopower, vs oxygen content, $y$, for different Ca families. Inset: calibration of holes $h$ per Cu(2)O$_2$ layer from thermopower, $S$, at $T=290$ K for the fully reduced samples. }

 \label{fig:2}
\end{figure}
%\end{widetext}

The figure shows the well known fact that oxygen does not contribute to hole transfer up to  $y_{t}\approx 0.12-0.15$ (red arrow), since below this threshold O concentration only locally charge-neutral Cu(1)OCu(1) dimers are formed \cite{Uimin:1992}, while above $y_{t}$ hole doping increases almost linearly with $y$, as oxygen ions start forming negatively charged trimers. This is true whatever the calcium content, which proves that the oxygen doping mechanism remains nearly the same, with a minor dependence of $y_t$ on $x$ (dashed line in Fig.~\ref{fig:2}). The samples ($x=0$, 0.05 and 0.08), which collapse on the same curve in Fig.~\ref{fig:1}, still show a large difference in their total RT mobile hole content $h(y)$ at  the critical oxygen concentration $y_c=0.3$ (blue arrow in Fig.~\ref{fig:2}) for the appearance of superconductivity. \footnote{A slight reduction of the $h(y)$ slope with increasing $x$ above $y_{t}$, has more probably to do with different mobilities of $h_O$ and $h_{Ca}$, rather than with the above mentioned trimer formation threshold.} 

We also plot in Fig.~\ref{fig:3} the reduced critical temperature $T_c/T_{c,max}$ versus $h$, ($T_{c,max}$ is the maximum transition temperature of each calcium series, at optimum doping \cite{Bernhard:1996}), showing beyond doubt that the onset of superconductivity does not fall on the same curve as a function of the total hole content $h$. Each series of samples at constant $x$ follows its own curve, contradicting the suggested {\em universal} parabolic relation \cite{Presland:1991} between $T_c$ and $h$, which is not the critical parameter in cuprate superconductivity, as it is often assumed.  A similar situation is evidenced by NMR\cite{Keren:2006} in \calabalacuo, where the NQR interaction shows that not all doped carriers contribute to the superconducting order parameter.

%\begin{widetext}
\begin{figure}[]
\includegraphics[width=0.45\textwidth,angle=0]{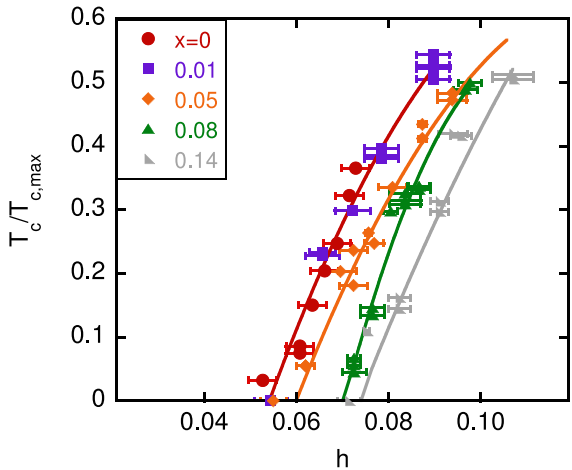}
\caption[ ]{ Scaling of transition temperatures with total holes: $T_c/T_{c,max}$ does not scale with total hole content, $h$. Here $T_{c,max}=93, 92, 91, 89$ and 87 K, respectively for $x=0, 0.01, 0.05, 0.08$ and 0.14 (from Ref.\onlinecite{Bernhard:1996}). }
 \label{fig:3}
\end{figure}
%\end{widetext}

The supercarrier pair density, $n_s$, was directly determined by transverse field (TF) $\mu$SR experiments. We measured two series with $x=0$, $x=0.05$, and variable $y$, plus two further samples ($x=0.08$, $y=0.43$ and $x=0.14$, $y=0.30$). The $\mu$SR experiment were performed on the MUSR spectrometer of the ISIS pulsed muon facility in the transverse field (TF) geometry\cite{Schenck:1986}, where an external magnetic field $\bm{H}$ is applied perpendicular to the initial muon spin $\bm{S}_\mu$ polarization. 
Field cooling the samples in $\mu_0 H=0.022$ T a flux lattice is formed, whose field inhomogeneity determines a depolarization rate of the muon spin precession, $\sigma(T)$, proportional to the inverse square of the London penetration depth, $\lambda_L$, hence to the supercarrier density $n_s$ \cite{Pumpin:1990}: $\sigma(T)\propto n_s(T)/m^*$ (where $m^*$ is the electron effective mass). All samples with $y<0.4$ display a coexistence of magnetism, below the freezing temperature $T_f$, and superconductivity, below $T_c$. The supercarrier density at zero temperature $\sigma_0=\sigma(T=0)$ must be obtained by extrapolating the data between these two transitions,  as in Ref.~\cite{Sanna:2004,Allodi:2006}.

%\begin{widetext}
\begin{figure}[]
\includegraphics[width=0.48\textwidth,angle=0]{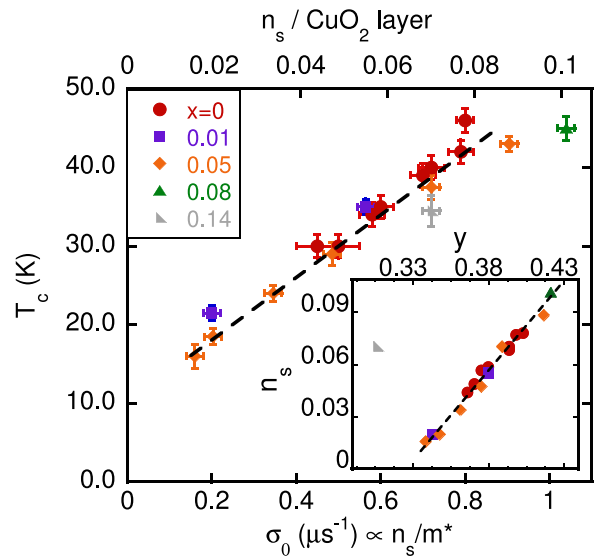}
\caption[ ]{ Scaling of transition temperatures with pair density: $T_c$ {\em does} scale with the $\mu$SR linewidth, $\sigma_0$, proportional to the supercarrier density, $n_s$. Inset: supercarrier density vs oxygen content. The samples for $0<x<0.08$ collapse on the same line which yields to $n_s\propto y$ (see text), i.e. only the oxygen hole fraction $h_O$ contribute to the supercarrier density for low calcium content.}
 \label{fig:4}
\end{figure}
%\end{widetext}

% RDR da qui in giù
The dependence of $T_c$ on $\sigma_0\propto n_s/m^*$, shown in Fig.~\ref{fig:4}, approaches the characteristic linear {\em Uemura-plot} behaviour \cite{Uemura:1989}. The upper axis of the plot represents the supercarrier density $n_s$ per CuO$_2$ plane, calculated in the clean limit approximation as $n_s[$hole/CuO$_2]\approx97500\cdot\sigma_0$, obtained by combining the relation $\sigma_0=7.58\cdot10^{-8}\lambda_L^{-2}$ in SI units\cite{Barford:1988}) with $\lambda_L^{-2}=\mu_0e^2n_s/m^*$. We assume a doping independent value of the effective mass, $m^*=3m_e$ \cite{Padilla:2005}. The plot of supercarrier density vs. oxygen content, $n_s(y)$, displayed in the inset of Fig.~\ref{fig:4}, shows that all samples with calcium content $0<x<0.08$ collapse on the same line, i.e. the dependence of $n_s$ on $y$ is linear. Since also holes injected by oxygen scale with $y$, $h_O\propto y-y_t$ (Fig.~\ref{fig:2}), the two linear relations imply that only the fraction of holes injected by oxygen, $h_O$, contributes to the supercarrier density for low calcium content.

Summarizing, our results show that in heavily underdoped compounds additional holes transferred from Ca participate to the RT normal metal behaviour (thermopower) with $h_{Ca}\propto x$. Their contribution however disappears from coherent conduction at low temperatures: whereas holes transferred from Cu(1)O chains contribute both to RT normal metal (above the threshold concentration $y_{t}\approx0.15$$, h_O\propto y-y_{t}$) and to superconducting properties (above the critical concentration $y_c=0.3$, $n_s \propto \sigma_0\propto Tc \approx k(y-y_{c})$), the Ca holes do not take part in the latter,  {\em as if} they were thermally activated. 

Notice that our results do not simply detect a reduction of coherent vs. metal hole densities: they furthermore indicate a mechanism of hole tagging, since the hole origin (i.e. whether they are transferred from Ca or from O) determines their behaviour. A direct explanation of this tagging may be that, as the perovskitic structure hosts insulating and metallic buffer layers, weakly interacting with each other only through the charge transfer mechanism, a further semiconducting component may come into play in \ybcoy. The interplay of semiconducting and metallic bands is however far from trivial when the latter is driven towards non Fermi liquid behaviour, as it is the case in underdoped cuprates. 

In particular it is difficult to reconcile this tagging with simple one-band Hubbard models,\cite{Zhang:1988}, and more realistic correlated-electron band structure calculations are called for. Whether these may be based on perturbative schemes (e.g. LDA+U \cite{Elfimov}) or on more sophisticated variations of the Hubbard model is an open question, but it is clear that material-specific predictions must consider multiple bands.

We further argue that our results have a consequence on the broader understanding of cuprates. The activated transport that we indirectly deduce in \ybcocay\ is a generic feature of the extremely underdoped regime in compounds where cation substitution is the only doping mechanism. For instance Hall effect measurements\cite{Ono:2007} identify similar activated mobile holes in \lsco.  Importantly, it was pointed out that the activation energy is proportional to $x^{-1}$ and that this matches also quantitatively the pseudogap features\cite{Gorkov:2006} determined by ARPES. Recent theoretical calculations\cite{Sboychakov:2007} show that a two-band Hubbard model may yield two distinct excitations that fit this pictures, providing a non trivial origin for the activated behaviour. 
% RDR
A straightforward conjecture is that {\em for all} cuprates the two features, namely the activated behaviour and the pseudogap excitations, are two sides of the same coin.  If we assume this conjecture to be valid for the \ybcocay\ system as well, our results suggest that pseudogap excitations are specifically related to Ca-tagged doping, a charge transfer mechanism which implies local Jahn-Teller distortion, i.e the presence of more than one band.
This, together with disorder and Coulomb scattering (inevitable with cation substitution) are likely to be essential ingredients for any realistic model of actual cuprates.

%\subsection{Acknowledgments}
%\begin{acknowledgments}
% put your acknowledgments here.
We acknowledge the support of PRIN06 ``Search for critical parameters in high $T_c$ cuprates'', partial support of PRIN05 ``Coexistence of magnetism and metallicity in high-Tc superconducting oxides.'', NMI3-Access and the staff of the ISIS facility (MUSR and GEM-XPRESS). We thank G. Calestani and L. Righi for help in diffraction refinements, G.~Guidi, G.~Allodi, A. Keren, A. Damascelli and V.~Fiorentini for fruitful discussions. 
%\end{acknowledgments}

%\bibliographystyle{papertry}
\bibliography{2bands}

\end{document}